\documentclass[prl,twocolumn]{revtex4}
\usepackage{amsfonts}
\usepackage{amsmath}
\usepackage{amssymb}
\usepackage{graphicx}

\newcommand{\BE}{\begin{equation}}
\newcommand{\BEA}{\begin{eqnarray}}
\newcommand{\EE}{\end{equation}}
\newcommand{\EEA}{\end{eqnarray}}

\newcommand\Tr{\mathrm{Tr}\,}
\newcommand\R\rangle
\renewcommand\L\langle
\renewcommand\r\right
\renewcommand\l\left
\newcommand\dd{\mathrm{d}}
\newcommand\ev{\mathbf{e}}
\newcommand\Sv{\mathbf{S}}
\newcommand\Kv{\mathbf{K}}

\newcommand\Mc{\mathfrak{M}}
\newcommand\mc{\mathfrak{m}}

\begin{document}
\title{The Hubbard model on the triangular lattice: Spiral order and spin liquid}
\author{Peyman Sahebsara}
\author{David S\'{e}n\'{e}chal}
\affiliation{D\'{e}partement de physique and Regroupement qu\'{e}b\'{e}cois sur les mat\'{e}riaux de pointe, Universit\'{e} de Sherbrooke, Sherbrooke, Qu\'{e}bec, Canada, J1K 2R1}
\date{November 2, 2007}

\begin{abstract}
We investigate the half-filled Hubbard model on an isotropic triangular lattice with the variational cluster approximation.
By decreasing the on-site repulsion $U$ (or equivalently increasing pressure) we go from a phase with long range, three-sublattice, spiral magnetic order, to a non-magnetic Mott insulating phase -- a spin liquid -- and then, for $U\lesssim6.7t$, to a metallic phase.
Clusters of sizes 3, 6 and 15 with open boundary conditions are used in these calculations, and an extrapolation to infinite size is argued to lead to a disordered phase at $U=8t$, but to a spiral order at $U\gtrsim 12$.
\end{abstract}

\maketitle


The effect of geometric frustration on quantum magnetism is still a very active field of investigation.
The quantum Heisenberg model on a two-dimensional square (bipartite) lattice exhibits long-range N\'eel order, but that order is suppressed on an isotropic triangular lattice.
In that case, the classical ground state is a spiral configuration in which the magnetization on each of the three sub-lattices is oriented at 120$^\circ$ of the other two.
For a while, it was conjectured that quantum fluctuations around that classical ground state would be strong enough to destroy this ordered pattern, but there is now a quasi-consensus that this is not the case \cite{Bernu,Capriotti}.
The latest Monte Carlo studies of the quantum Heisenberg model on a triangular lattice point towards a sub-lattice magnetization of $\mc\approx0.41$ in the ground state \cite{Capriotti}.

However, real antiferromagnets are better described by the Hubbard model,
\BE\label{eq:Hubbard}
H=- t\sum_{\langle ij\rangle,\sigma}c^\dagger_{i\sigma}c_{j\sigma}+ U \sum_i n_{i\uparrow} n_{i\downarrow}
\EE
where $t$ is the hopping amplitude between neighboring sites, $c_{i\sigma}$ 
destroys an electron of spin $\sigma$ at site $i$ and $U$ is the on-site Coulomb repulsion.
The Heisenberg model is recovered in the strong coupling limit ($U\gg t$), with direct-exchange constant $J\sim 4t^2/U$.
Finite-$U$ effects are potentially important on real systems to which the Heisenberg model is usually applied.
Such effects are often incorporated as ring-exchange terms in spin models~\cite{Motrunich:2005}, but their origin can be traced back to the Hubbard model itself~\cite{Delannoy:2005}.
For instance, the organic conductor $\kappa$-(BEDT-TTF)$_2$Cu$_2$(CN)$_3$ may be described by a Hubbard model on an almost isotropic and half-filled triangular lattice~\cite{Mckenzie}, and this material is conjectured to be in a spin liquid (i.e. magnetically disordered and insulating) phase \cite{Shimizu:2003}.
So is the triangular antiferromagnet $\rm EtMe_3Sb[Pd(dmit)_2]_2$~\cite{Itou:2007}.
The question that arises in this case is whether such a state is compatible with a Hubbard model description.
In this paper, we will argue that it is, i.e., that the Hubbard model on a triangular lattice exhibits a spin liquid phase at intermediate values of $U$ (e.g. $U\sim 8$) although it exhibits spiral magnetic order at stronger coupling (e.g. at $U=12$).

The Hubbard model on an {\it anisotropic} triangular lattice has been studied by various methods.
The 120$^{\circ}$ spiral state has been studied in the mean-field approximation \cite{Cote:1995,Singh}, and a spin stiffness analysis points to a loss of order for $U\lesssim6$ \cite{Singh}.
In the isotropic case, slave-bosons methods were used to obtain a phase diagram qualitatively similar to the Hartree-Fock results~\cite{Capone}, with a transition from a metallic phase to a magnetic phase, with no intercalated spin liquid phase.
On the other hand, the presence of a Mott phase was confirmed in Refs.~\onlinecite{Morita:2002, Sahebsara:2006, Kyung:2006, Aryanpour:2006}, however without confronting it with a spiral magnetic order.

In this work we use the (zero temperature) variational cluster approximation (VCA) \cite{Potthoff:2003}.
This method goes beyond mean field and takes into account exactly the effects of strong short-range correlations.
As $U$ is increased, we show that the system goes from a metallic phase to a non magnetic, insulating phase (i.e., a spin liquid) at around $U\approx 6.7$, and then to a magnetic, spiral phase at larger values of $U$.
Our treatment involves exact solutions of the model on triangular clusters of 3, 6 and 15 sites, as well as an extrapolation to infinite size.

\begin{figure}[tbp]
\centerline{\includegraphics[width=7.5cm]{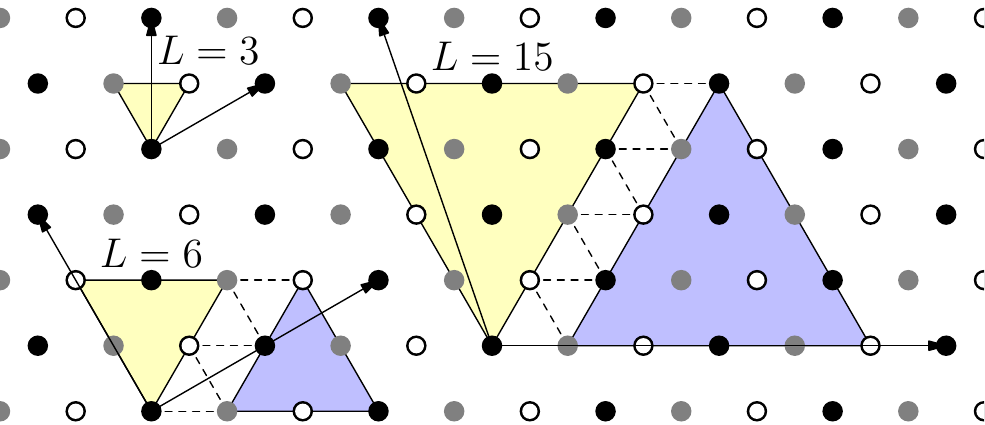}}
\caption{(Color online) clusters used in our study.
The 6-site and 15-site clusters tile the lattice only when paired with identical, inverted clusters.
Superlattice basis vectors are shown.}
\label{fig_clusters}
\end{figure}

\paragraph{The Variational Cluster Approximation.}
The VCA \cite{Potthoff:2003b} is a quantum cluster approach to the Hubbard model that rests on Potthoff's self-energy functional approach (SFA) \cite{Potthoff:2003}.
It has been applied, for instance, to the problem of competing phases in the high-$T_c$ cuprates \cite{Senechal:2005} and in the layered organic conductors \cite{Sahebsara:2006}.
The SFA involves a functional $\Omega_{\bf t}[\Sigma]$ of the self-energy, parametrized by the one-body terms collectively labeled by $\bf t$, that is stationary at the physical self-energy of the system: $\delta\Omega/\delta\Sigma=0$.
The SFA introduces a {\it reference} Hamiltonian $H'$, with the same two-body interaction as the original Hamiltonian $H$, but with a different one-body part, so that $H'$ may be solved numeri\-cally.
The functional $\Omega_{\bf t}[\Sigma]$ is then 
\BE\label{eq:omega1}
\Omega_{\bf t}[\Sigma]=\Omega_{\bf t'}[\Sigma] - \Tr\ln\l( G_0^{-1}\kern-0.15em-\kern-0.15em\Sigma\r) 
+ \Tr\ln\l( G'_0{}^{-1}\kern-0.15em-\kern-0.15em\Sigma\r)
\EE
where $G_0$ and ${G^\prime}_0$ are the non-interacting Green functions of $H$ and $H^\prime$, respectively.
At the physical self-energy, this functional is the grand potential $\Omega$. 

In VCA, $H'$ is obtained from $H$ by (i) tiling the lattice into a super-lattice of identical clusters, and removing all inter-cluster hopping terms and (ii) introducing on the clusters Weiss fields that allow for broken symmetry phases.
Then the Weiss fields (collectively denoted $h$ in what follows) are used as variational parameters and the functional $\Omega_{\bf t}[\Sigma]$ reduces to a function $\Omega_{\bf t}(h)$ given by
\BE\label{eq:omega2}
\Omega_{\bf t}({\bf t}')=\Omega'\kern-0.1em - \kern-0.1em\int_C \frac{d\omega}{2\pi}\sum_{\Kv}\ln\det\left(
1 \kern-0.1em + \kern-0.1em (G_0^{-1}\kern-0.2em -G_0'{}^{-1})G'\right)
\EE
where $G^\prime$ is the exact Green function of $H^\prime$, $\Omega^\prime$ the exact grand potential of $H^\prime$ and the sum is over wave-vectors $\Kv$ of the Brillouin zone of the super-lattice.
In practice, one searches for the stationary points of the above function, whose evaluation requires, at each point $h$, the exact solution of the Hamiltonian $H'$ defined on a finite-size cluster.
At these points, the self-energy $\Sigma$ of $H'$ is considered an approximation to the physical self-energy and is used to construct the Green function $G = (G_0^{-1} - \Sigma)^{-1}$ of the lattice model.
Thus, VCA provides us with an approximate Green function of the system, allowing the calculation of spectral and thermodynamic properties, both in broken symmetry phases and normal phases.

In investigating broken symmetry phases, VCA is superior to static mean field approaches in that it does not require any factorization of the interaction, and short-range correlations (within a cluster) are taken into account exactly. The Green function obtained is still defined on the infinite lattice. The only approximation comes from the limited space of self-energies on which the variational principle is applied, limited by the cluster size and by the number of variational parameters used.

\begin{figure}[tbp]
\centerline{\includegraphics[width=7.5cm]{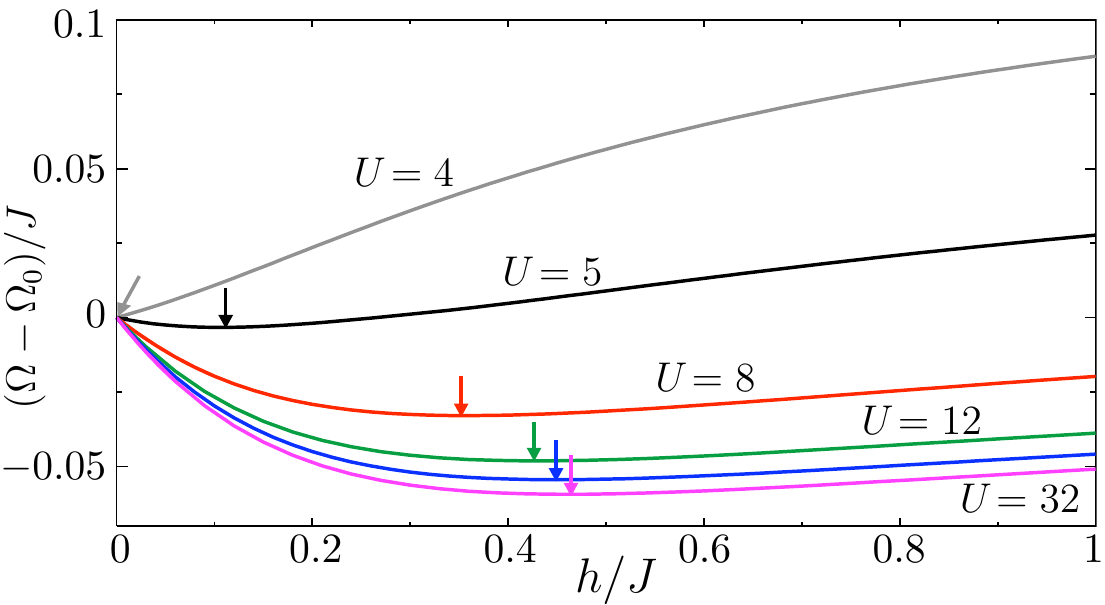}}
\caption{(color online) Scaled Potthoff functional $\Omega$ as a function of Weiss field $h$ for various values of $U$ (3-site cluster).
The local minima are indicated by arrows.}
\label{fig_omega_Weiss}
\end{figure}

\paragraph{Clusters for the spiral order.}
The Weiss field term associated with the spiral magnetic order may be expressed as $H'_h = h\hat\Mc$, where
\BE\label{eq:weiss}
\hat\Mc =  \l\{ \sum_{i\in A} \ev_A\cdot\Sv_i
+\sum_{i\in B} \ev_B\cdot\Sv_i
+\sum_{i\in C} \ev_C\cdot\Sv_i \r\}
\EE
where $A$, $B$ and $C$ stand for the three sub-lattices of the triangular lattice, as shown on Fig.~1 by different shades of gray.
The unit vectors $\ev_{A,B,C}$ are oriented at 120$^\circ$ of each other, and the spin operator is
$\Sv_i = c^\dagger_{i,\alpha}\mathbf{\sigma}_{\alpha\beta}c_{i,\beta}$.

The clusters used in applying VCA to the triangular lattice are depicted on Fig.~1.
They all have a triangular shape and treat the three sub-lattices on the same footing.
Since the $L=6$ and $L=15$ clusters do not tile the lattice by themselves, they are paired with their rotated mirror-image to define a true super-lattice, in the Bravais sense.
More explicitly, the Green function $G'$ of the super-lattice's unit cell (the union of the cluster and of its mirror-image) is given by
\BE
G'{}^{-1} = G_1'{}^{-1} + G_2'{}^{-1} + t_{12}
\EE
where $G_1'$ is the Green function of the cluster itself (site and spin indices suppressed), $G_2'$ that of its rotated mirror image (a simple transformation of $G_1'$) and $t_{12}$ the hopping matrix linking the two (dashed links on Fig.~\ref{fig_clusters}).

The variational parameters used in this work are the Weiss field $h$ multiplying $\Mc$ (see Eq.~(\ref{eq:weiss})) and the chemical potential $\mu'$ of the cluster.
Treating $\mu'$ as a variational parameter instead of setting $\mu'=\mu$ ensures thermodynamic consistency, i.e., that the densities obtained by calculating $n=\Tr G$ and $n=-\partial\Omega/\partial\mu$ coincide.
Fig.~\ref{fig_omega_Weiss} illustrates the $h$ dependence of $\Omega_{\bf t}(h,\mu')-\Omega_{\bf t}(0,\mu')$ for the value of $\mu'$ corresponding to the solution, for several values of $U$ and on a 3-site cluster.
As one can see, a local minimum exists as a function of $h$ for $U\geq5$.
$h$ and $\Omega_{\bf t}$ were divided by $J=4t^2/U$ in order to emphasize the strong-coupling scaling behavior.

\begin{figure}[tbp]
\centerline{\includegraphics[width=8.5cm]{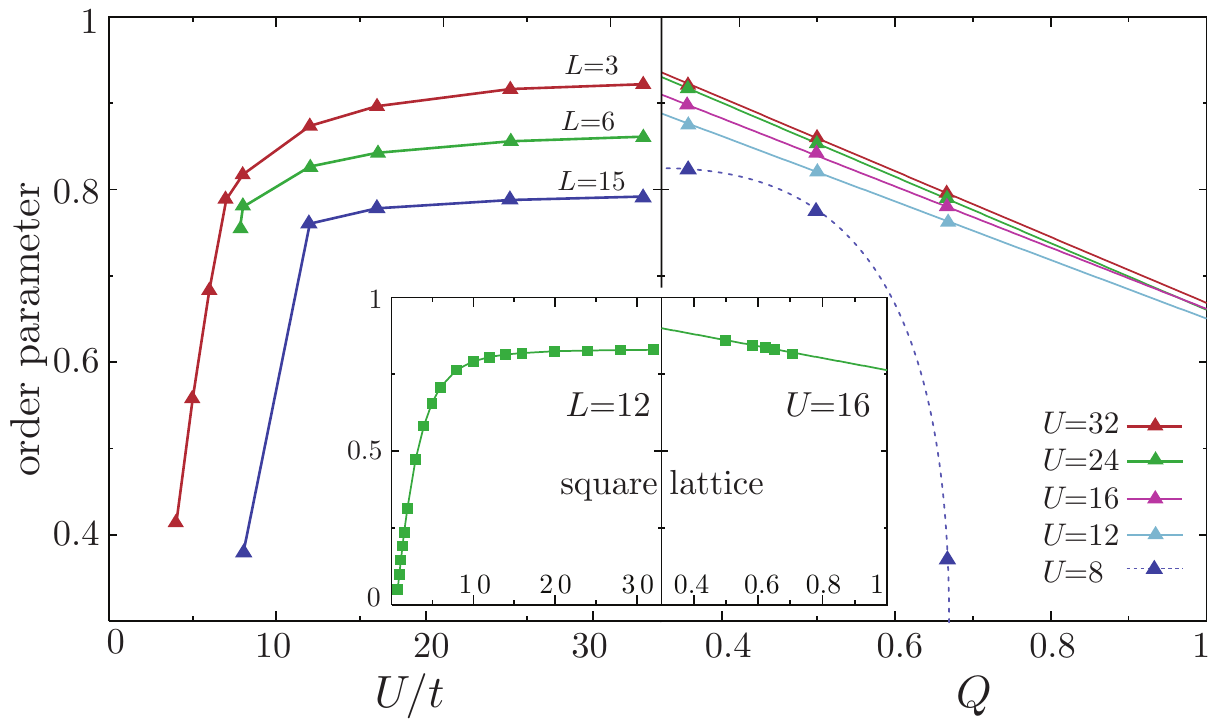}}
\caption{(color online) Left panel: $U$ dependence of the spiral order parameter for $L=3$, 6 and 15.
Right panel: Spiral order parameter as a function of scaling parameter $Q$, for various $U$'s. 
The $U=8$ curve is a guide to the eye only.
Left inset: N\'eel order parameter as a function of $U$ on a square lattice (12-sites).
Right inset: The same as a function of $Q$.}
\label{fig_OP}
\end{figure}

The Newton-Raphson algorithm is used to locate the values of $h$ and $\mu'$ that make the function $\Omega_{\bf t}(h,\mu')$ (Eq.~(\ref{eq:omega2})) stationary.
The self-energy obtained at that point is then used to construct the approximate lattice Green function.
The spiral order parameter $\mc$, i.e., the expectation value $\langle\hat\Mc\rangle$ divided by the number of lattice sites, is calculated from that Green function as
\BE\label{OP}
\mc=2i \int \frac{\dd^2K}{(2\pi)^2} \int \dfrac{\dd\omega}{2\pi} G_{ab}(i\omega,\Kv) \Mc_{ba} 
\EE
where the frequency integral is taken along the positive imaginary axis, and the sum over repeated indices is implicit.
$a,b$ are composite indices including both cluster site and spin: $a \equiv (i,\sigma)$.
$\Mc_{ab}$ is a matrix of real numbers expressing $\hat\Mc$ as a one-body operator:
$\Mc_{ab}c^\dagger_a c_b$.

The calculation is performed for several values of the lattice chemical potential $\mu$ until the density $n$ is close enough to half filling ($n=1$).
In practice, this is easily accomplished when $U$ is large enough for a spectral gap to open ($U\approx6$ and above).

The left panel of Fig.~\ref{fig_OP} shows the spiral order parameter as a function of interaction strength $U/t$, for fixed cluster size.
The order parameter is seen, as expected, to saturate at strong coupling.
The transition from the magnetic to the disordered state seems of first order, but the discontinuity depends on cluster size and might disappear in the thermodynamic limit; we could perform no quantitative analysis on this matter.

The values found for the Weiss field $h$ and the order parameter $\mc$ depend both on $U$ and on cluster size.
We are naturally interested in the infinite-size extrapolation of these, since an ordered solution found on a finite cluster can disappear in the thermodynamic limit because of long wavelength fluctuations of the order parameter.
Such an extrapolation is very difficult to do with the small clusters at our disposal.
Moreover, these clusters have open, not periodic, boundary conditions.
This implies that the number of sites of the cluster ($L$) is not the only scaling parameter: the size of its boundary could also be significant.
We define a scaling parameter $Q$ as the number of links within the cluster divided by the total number of links of the original lattice within a unit-cell of the super-lattice of clusters.
$Q$ increases with cluster size and reaches unity in the thermodynamic limit.
It is equal to $1/3$, $1/2$, and $2/3$ respectively for the 3-, 6-, and 15-site triangular clusters.

\begin{figure}[tbp]
\centerline{\includegraphics[width=\hsize]{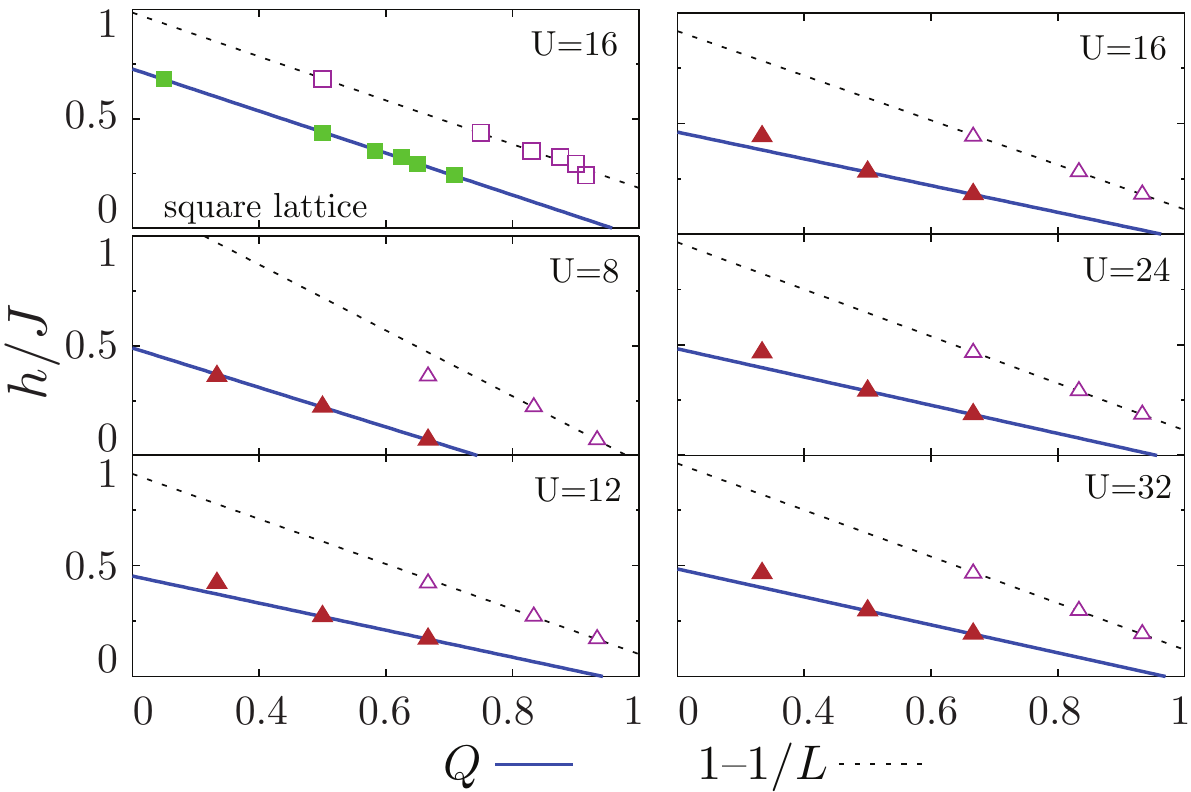}}
\caption{(color online) Scaled Weiss field as a function of $Q$ (solid lines) and $1-1/L$ (dashed lines) for various values of $U$. 
The data are obtained for 3- 6- and 15-sites triangular clusters.
Top left panel: square lattice results at $U=16$ for the N\'eel Weiss field, with $L=$ 2, 4, 8, 10, 12 and 16 sites.}
\label{fig_extrapol}
\end{figure}

We expect the Weiss field to vanish in the thermodynamic limit, as it is then no longer necessary to stabilize order.
If the Weiss field extrapolates to zero at $Q<1$, this is to be interpreted has a suppression of order due to long wavelength fluctuations.
Fig.~\ref{fig_extrapol} displays the spiral Weiss field as a function of both $Q$ and $1-1/L$ for several values of $U$, as well as the N\'eel Weiss field for several square lattice clusters at $U=16$ and half-filling.
The square-lattice results show that $Q$ is a better scaling parameter than $1-1/L$ since the values of the N\'eel Weiss field neatly fall on a straight line. This line crosses the abscissa very close to $Q=1$, as it should since long-range N\'eel order is expected in the square-lattice case.
In the triangular case, however, the 3-site cluster is too small to be in the scaling regime, and we must rely only on the 6- and 15-site clusters to extrapolate towards $Q=1$.
Fig.~\ref{fig_extrapol} shows that the Weiss field extrapolates to zero very near $Q=1$ for all values of $U$ studied except $U=8$.
Even though $1/L$ scaling looks superficially better for triangular clusters, it extrapolates beyond $1/L=0$ (except for $U=8$).
Whatever the extrapolation scheme, we conclude that there is no long-range order at $U=8$.
Thus long-range spiral order is established somewhere between $U=8$ and $U=12$.

\begin{figure}[t]
\centerline{\includegraphics[width=8.5cm]{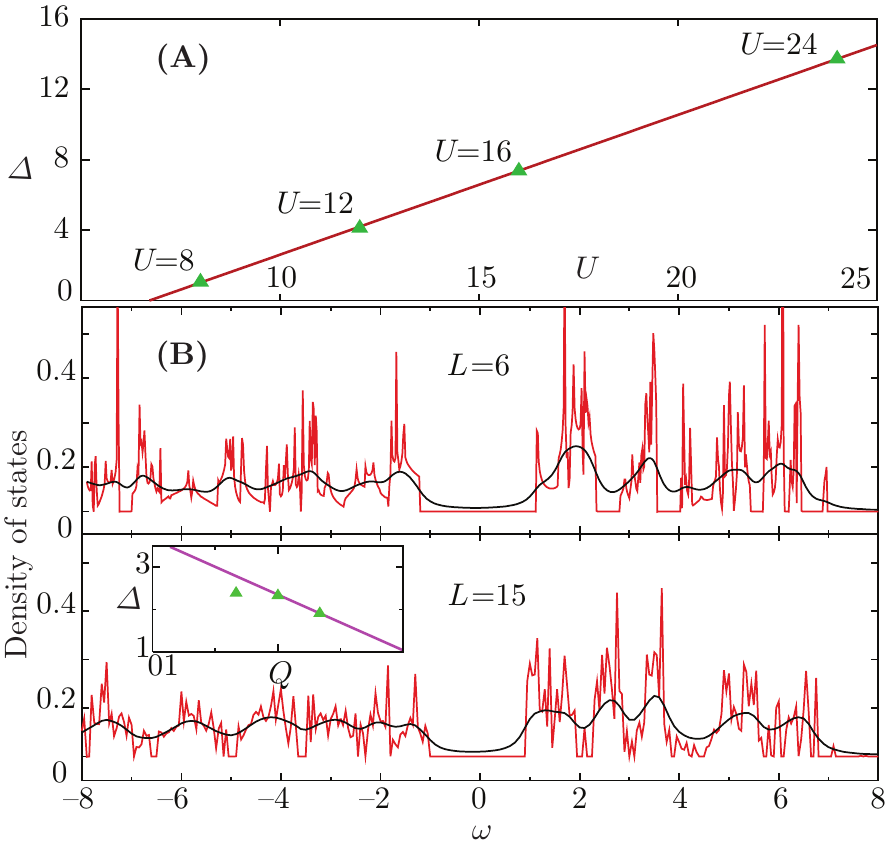}}
\caption{(color online) Bottom panel (B) : Density of states at $U=8$ (see text for details).
Inset : the Mott gap as a function of $Q$. Top panel (A) : $U$ dependence of the infinite-size extrapolated Mott gap.}
\label{fig_dos}
\end{figure}

The right panel of Fig.~\ref{fig_OP} shows the spiral order parameter $\mc$ as a function of the scaling parameter $Q$. 
A linear extrapolation to $Q=1$ yields $\mc\sim0.65$, which is larger than values obtained for the Heisenberg model by Monte Carlo methods \cite{Capriotti}.
Thus, despite the extrapolation, VCA still exaggerates the tendency of the system to order.
Indeed, a similar analysis for the square lattice N\'eel order (inset of Fig.~\ref{fig_OP}) yields an extrapolated magnetization of 0.74, whereas the accepted value is closer to 0.61\cite{Sorella:1998}.
Therefore, if VCA predicts a magnetically \textit{disordered} state, it is very likely correct.
Incidently, it is impossible to extrapolate the $U=8$ order parameter to infinite size, which is a further indication of the absence of order at $U=8$. 

We now turn our attention to the Mott transition.
Fig.~\ref{fig_dos}B shows the density of states (DOS), calculated by integrating the lattice Green function $G(\omega,\Kv)$ over wave-vectors, for $U=8$.
The smooth curve is obtained by giving the complex frequency an imaginary part $\eta=0.2t$, equivalent to a Lorentzian broadening of the spectral function delta peaks.
The red (jagged) curve is the point-wise extrapolation towards $\eta\to0$ of the DOS calculated at $\eta=0.01$, 0.005 and 0.002.
This extrapolation allows for a better numerical estimate of $\Delta$.
This estimate can be extrapolated to infinite size (with the help of the scaling parameter $Q$, see inset).
The extrapolated values can be plotted as a function of $U$ to extract a critical value $U_c$ for the Mott transition (Fig~\ref{fig_dos}A).
We find $U_c\approx 6.7$.

To conclude, the system is predicted to be a metal for $U\lesssim 6$, a magnetically disordered Mott insulator (or spin liquid) at intermediate values of $U$, and a spiral magnet at larger values of $U$ (already at $U=12$).
A coexistence of the latter two phases could occur if the transition between the two were still of first order in the thermodynamic limit, although we cannot conclude on the matter.
This is in contrast to the square-lattice model, in which the Mott transition is pre-empted by N\'eel order down to $U=0$ (inset of Fig.~\ref{fig_OP}).

Discussions with A.-M.~S.~Tremblay are gratefully acknowledged, as is
support by NSERC (Canada).
The computational resources were provided by the R\'eseau qu\'eb\'ecois de calcul de haute performance (RQCHP).


\end{document}